\documentclass{emulateapj}

\shorttitle{Shocks in NGC 5929}
\shortauthors{Rosario et al.}

\newcommand{\kms}{km s$^{-1}$}

\begin{document}

\title{The Radio Jet Interaction in NGC 5929: Direct Detection of Shocked Gas}

\author{D.J. Rosario\altaffilmark{1}, M. Whittle\altaffilmark{2}, 
C.H. Nelson\altaffilmark{3}, A.S. Wilson\altaffilmark{4}}

\altaffiltext{1}{Astronomy Department, University of California, Santa Cruz,
CA 95064; rosario@ucolick.org}

\altaffiltext{2}{Astronomy Department, University of Virginia, Charlottesville,
VA 22903; dmw8f@virginia.edu}

\altaffiltext{3}{Physics and Astronomy, Drake University, Des Moines, 
IA 50311--4505; charles.nelson@drake.edu}

\altaffiltext{4}{Astronomy Department, University of Maryland, College Park,
MD 20742; deceased}

\begin{abstract}

We report the discovery of kinematic 
shock signatures associated with a localized radio jet interaction in the 
merging Seyfert galaxy NGC 5929. We explore the velocity-dependent ionization 
structure of the gas and find that low ionization gas at the interaction site 
is significantly more disturbed than high ionization gas, which we
attribute to a local enhancement of shock ionization due to the influence of the jet. 
The characteristic width of the broad low-ionization emission 
is consistent with shock  velocities predicted from the ionization conditions of the gas.
We interpret the relative prominence of shocks to the high density of gas in nuclear
environment of the galaxy and place some constraints of their importance 
as feedback mechanisms in Seyferts.

\end{abstract}


\section{Introduction} \label{sec1}

Most Seyfert AGN are associated with weak
nuclear radio sources which show radio spectral indices and morphologies
(if resolved) consistent with synchrotron-emitting jets and lobes 
\citep{ulv84, nagar99}, though their physical properties are poorly constrained.
However, they are effective avenues of kinetic and thermal feedback from the active nucleus
and may play an important role in determining the evolution of the central spheroid. 

In the small fraction of Seyferts with kpc-scale radio jets, several studies have uncovered 
clear signatures of interactions between the jet and surrounding gas, in the form of disturbed
emission line profiles in the inner Narrow-Line Region (NLR), as well as close
associations between resolved jet structures and NLR gas \citep[e.g.][]{whittle88,capetti96,fws98,cooke00,cecil02,whittle04}.
Depending on the physical make-up of the jet, relativistic or ram pressure can drive 
fast shocks, compressing, sweeping up and altering the appearance 
of the NLR. Postshock gas, with a temperature of several $10^7$ K, is
a source of ionizing photons which couple the shock properties to the ionization
state of the surrounding gas \citep[e.g.][]{ds96}. Shocks and winds can also 
change the distribution of ISM phases in the NLR by destroying and ablating 
clouds \citep{fragile05}. While the active nucleus is usually the dominant source of 
ionization even in strongly jetted Seyferts \citep[e.g.][]{whittle05}, widespread shocks 
driven by a jet can alter the ionization of the NLR by affecting the properties of the ISM.

Studies of shock structure and energetics \citep{ds96, allen08} predict strong differences 
between the emission line spectrum of dense post-shock gas and gas
that is ionized by,  but not in direct dynamical contact with, the shock (the precursor). 
Therefore, a clear signature of shock ionized gas is a difference between the line profiles of
low and high ionization lines, which are preferentially produced by post-shock and
precursor gas, respectively \citep{whittle05}.  

In this work, we present an HST/STIS spectroscopic study of NGC 5929, a local Seyfert
galaxy with a well-studied bi-polar radio jet. Previous ground-based spectroscopic 
studies find evidence of a localized interaction between the jet and the 
near-nuclear emission line gas \citep{whittle86,wakamatsu88,taylor89,wilson89,ferruit97}. 
The datasets and analysis methods used are briefly reviewed 
in $\S2$. Direct shock features and a picture of the interaction are developed in 
$\S3$ and $\S4$. We discuss the role of shocks in Seyferts and AGN feedback in $\S5$.
NGC 5929 has a systemic heliocentric velocity of $cz\,=\,2492$ km s$^{-1}$ 
based on the stellar aborption line redshift of \citet{nelsonnwhittle95}, which corresponds to  
$161$ pc arcsec$^{-1}$  (H$_0 = 75$ km s$^{-1}$ Mpc$^{-1}$).

\begin{figure*}[ht]
\label{corrplot}
\centering 
\includegraphics[width=0.7\columnwidth,angle=270]{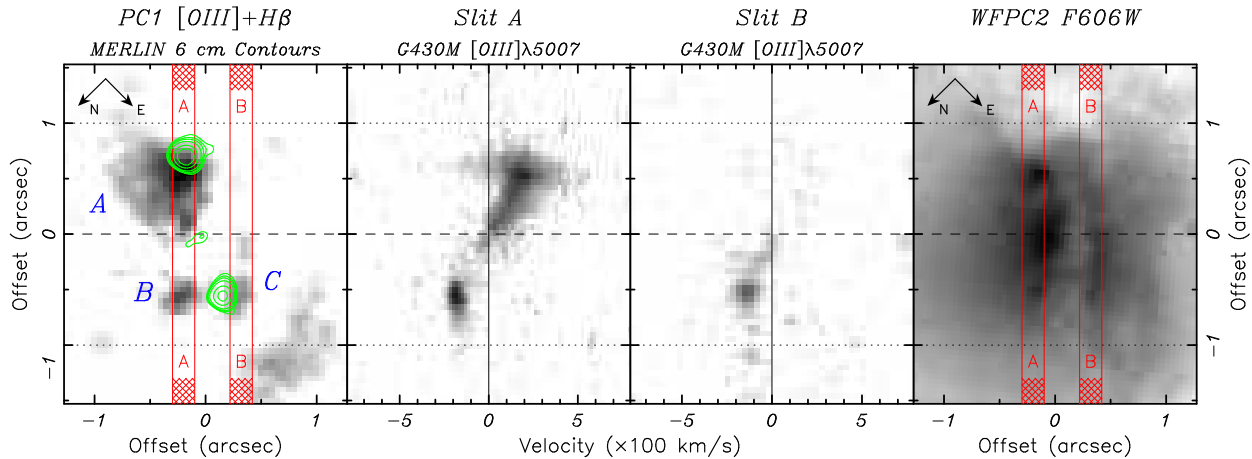}
\caption[NGC 5929: Combined datasets]
{ Panels of images and spectra of the nuclear region of NGC 5929. The 2D spectra in Panels 2 and 3
correspond to STIS long-slit apertures A and B respectively, as indicated on the images plotted in 
Panels 1 and 4. The radio map (gaussian smoothed by $0\farcs3$ to bring out its structure)
is plotted in Panel 1 in contours.
}
\end{figure*}
  
\section{Observations and Reductions} \label{sec2}

\subsection{STIS Dataset} \label{sec2p1}

   Our principal dataset is two  medium-dispersion STIS G430M long-slit
spectra covering the lines of the [\ion{O}{3}]$\lambda \lambda 5007, 4959$ 
doublet and H$\beta$ (HST Program: GO 8253; PI : Whittle). 
The data was spectrophotometrically calibrated using the CALSTIS pipeline.
For each spatial increment, the continuum was modeled using a lower 
order polynomial and subtracted.
These spectra were combined to generate a continuum-free 2-D spectrum
from both STIS slits
The equivalent width of H$\beta$ is always greater than 10 \AA\ in the NLR of NGC 5929, 
and $\approx 140$ \AA\ at the location of the interaction. Therefore,
corrections to the emission line strengths and profiles 
due to Balmer absorption are negligible.

   
\begin{deluxetable}{cccc}
\tabletypesize{\scriptsize}
\tablewidth{\columnwidth}
\tablecaption{STIS observations \tablenotemark{\dag}} \label{tab1}
 
\tablenotetext{\dag}{Date: 07/02/00; Grating: G430M;  $\Delta\lambda$ (\AA\ pix$^{-1}$): 0.277; PA: $-134\arcdeg6$}
\tablenum{1}
\tablehead{\colhead{Aperture}&\colhead{Dataset}&\colhead{Exp. (s)}&\colhead{Nuc. Offset (")} }
\startdata
NGC 5929 A & O5G403010 & 1524 & 0.198 \\
NGC 5929 B & O5G403020 & 600 & 0.390 \\
\enddata
\end{deluxetable}

\subsection{HST and Radio Images} \label{sec2p2}

  Emission line maps of NGC 5929 in 
[OIII]$\lambda\lambda4959,5007$+H$\beta$  were prepared from 
archival HST narrow/medium-band WF/PC-1 images 
(Program: GO 3724; PI: Wilson). Details of these maps can be found in  
\citet{bow94}.  
In addition, a high S/N F606W WFPC2 archival image of the galaxy 
was used to trace the stellar and dust geometry in and around the NLR.
A reduced and calibrated 5 GHz radio map of NGC 5929 was obtained
from the MERLIN archive (http://www.merlin.ac.uk/archive/). This data was first presented in \citet{su96}.

The various HST datasets were registered to a common astrometric frame using image
cross-correlation techniques. The dust lane in 
NGC 5929 crosses the NLR, effectively obscuring the true position of the nucleus. 
\citet{latt97} use the astrometric information of stars in the WF/PC-1 images of the galaxy
to compare the locations of the continuum peak and the unresolved core 
in a MERLIN radio map. Assuming the radio core corresponds to the true nucleus, 
they find that the WF/PC-1 continuum peak is offset by $0\farcs1 \pm 0\farcs05$ NW
from the radio core. We have included this small correction in our final 
registered images (see Panel 4 in Figure \ref{corrplot}
for a sense of the amount of correction involved). 

\section{The NLR of NGC 5929}  \label{sec3}

\subsection{Descriptive Framework} \label{sec3p2}

NGC 5929 is part of a major galaxy merger with NGC 5930 (Arp 90). Clear signs of tidal 
tails and disturbed gaseous kinematics are seen in ground-based 
two-dimensional spectra of the merger system \citep{lewis93}, and 
twisting, filamentary dust structure is visible in the nuclear region 
(Panel 4 of Fig. \ref{corrplot}) 

The radio source has been imaged at high resolution
with the VLA \citep{ulv84,wilson89} and MERLIN \citep{su96}. 
It exhibits a triple structure, with two bright compact steep-spectrum
hotspots on either side of a faint flat-spectrum core that is unresolved
even in the highest resolution maps. Low surface brightness
emission joins the hotspots to the radio core along  PA $61^{\circ}$,
which is also the rough orientation of the elongated emission line distribution.  

The diverse imaging and spectroscopic datasets are showcased 
together in Figure \ref{corrplot}. In Panel 1, the [\ion{O}{3}]$+$H$\beta$
emission line image is displayed in greyscale, with the
STIS slit positions and radio map contours overplotted. Three emission line regions,
identified by \citet{bow94} are labelled A, B and C. The SW radio hotspot
coincides with the edge of Region A, while the NW radio hotspot lies near
Region B. From the broadband image in Panel 4, the emission
line features are revealed as parts of a possibly contiguous region,
crossed by a dust lane that obscures the line and continuum emission
and modulates the appearance of the NLR. This dust structure is part of
a filamentary network extending to several kpc \citep{malkan98}.
 
The [OIII]$\lambda 5007$ line from the two medium-resolution STIS 
spectra are plotted in Panels 2 \& 3 of Figure \ref{corrplot}. The slits
are oriented NE-SW from the bottom to the top of the panel, and a
dashed horizontal line marks the reference position on the slit, i.e, 
the point along the slit closest to the nucleus of the galaxy. Solid vertical 
lines indicate the systemic velocity. 

The [O III] line from both STIS spectra shows clear velocity
structure. A detailed treatment of the kinematics of the NLR  
is not the aim of this Letter and will be addressed in a later paper. 
A brief description will suffice here.
Slit A traverses the brighter emission line regions and therefore
gives the best picture of the ionized gas kinematics. The velocity
of the line peak increases along this slit from almost systemic
at the reference position to a maximum value of $+185$ \kms
at a nuclear radius of $\sim 0\farcs65$ SW (104 pc). This broad gradient
is mirrored, though less clearly, to the NE in both Slits
A and B. The gradient between the nucleus and Regions
B and C are similar in both slits, implying that both Regions 
are part of the same gaseous complex, bisected
in projection by the dust lane. The FWHM of the [O III] line goes through
a rapid transition from $\sim 125$ \kms\ to greater
than $200$ \kms\ at a nuclear distance of $0\farcs5$ in the 
brightest portion of Region A. Regions B and C, on the other hand,
exhibit uniformly narrow line profiles. 

\begin{figure}[t]
\label{ratiomap}
\centering 
\includegraphics[width=0.6\columnwidth,angle=270]{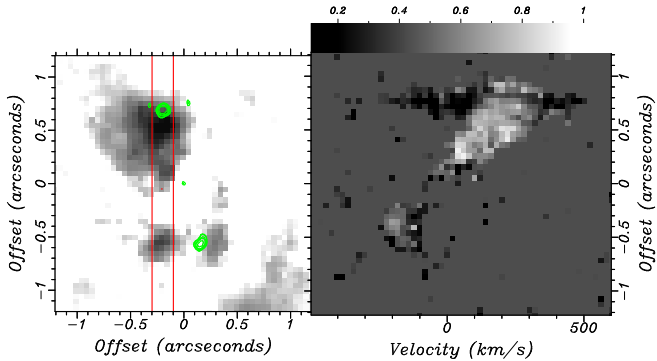}
\caption[NGC 5929: Overlay and Excitation ratio map]
{The [OIII] image with MERLIN 6 cm contours overlayed (left) 
and the [OIII]/H$\beta$ log ratio map (right). The location of the slit 
aperture is indicated on the image as a rectangle, corresponding to 
the spatial range of the ratio map. A grayscale lookup table is 
plotted in the bar above the map. A fiducial value of 0.43 is used to 
mask regions with no significant line emission, which helps to improve 
the visibility of the ratio map. 
}
\end{figure}

\subsection{Ionization Conditions} \label{sec3p3}

In Figure 2, a map of the [OIII]$\lambda5007$/H$\beta$ line
ratio from the Slit A spectrum is plotted, alongside the WF/PC-1 [O III] image, 
overlayed with the contours of the 5 GHz radio map
and the boundaries of the STIS aperture.  The ratio image was created by
first rebinning the two-dimension spectrum
of each line onto a common velocity range and then dividing them to
generate a map of the [O III]/H$\beta$ ratio
as a function of velocity and slit position.  This ratio of is quite sensitive to 
ionization state, yet insensitive to dust reddening.

\begin{figure}[t]
\label{profilecomps}
\centering 
\includegraphics[width=1.1\columnwidth]{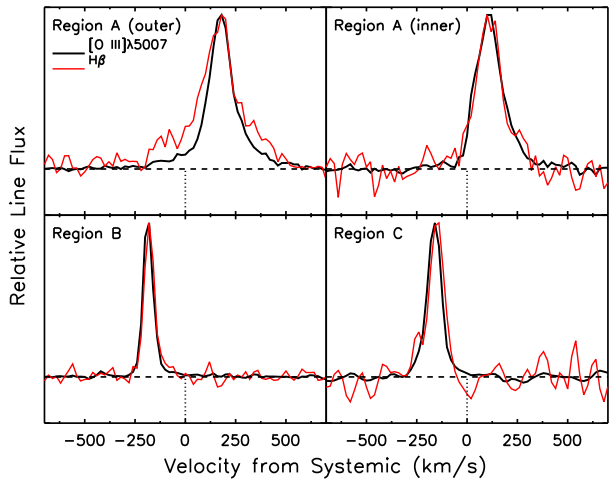}
\caption[NGC 5929: Profile Comparisons]
{A comparison of the line profiles of the [O III] line (black) and
H$\beta$ line (red) in four parts of the NGC 5929 NLR. The lines have been scaled
to a common peak value. The jet interaction, in the outer part of Region A, 
is associated with broad H$\beta$, unlike any other part of the emission line region.
}
\end{figure}

The [O III]$\lambda5007$/H$\beta$ ratio is in the range of $2-6$, intermediate between 
Seyferts and LINERs \citep{kewley06}. Variations in ionization are seen 
across the NLR. The average excitation of Region B is towards the low
end of the observed range, with average [O III]$\lambda5007$/H$\beta \sim 3$, 
while the inner part of Region A is more highly ionized ([O III]$\lambda5007$/H$\beta 
\sim 5$). However, surrounding the position of the bright knot in Region A, a sharp 
drop in the ratio is seen. Interestingly, this change in ionization is velocity
dependent. In the high velocity wings of the lines, the [O III]/H$\beta$ 
drops to almost unity implying a very low ionization state, while the central core of 
the lines remain at a more modest ionization. This trend is different from most
Seyferts where the high velocity gas tends to be of equal or \emph{higher} ionization
than low velocity material \citep[e.g.,][]{pelat81}. 

This difference between the average ionization of the NLR and Region
A is brought out well by comparing the velocity profiles of [O III]$\lambda 5007$
and H$\beta$ in different parts of the NLR, as is done in Figure 3.
The width of the H$\beta$ line compared to [\ion{O}{3}] is substantially higher 
in Region A, compared to both the inner part of the same Region and the profiles 
in Region B and C.

What is the reason for this peculiar ionization behaviour? We believe this
results from strong shocks driven by the radio plasma into the ISM. 
The compressed gas in the post-shock cooling zone is expected to share the 
high velocities of the shock, yet is predicted to have [O III]/H$\beta$ in 
the range of $0.7-4.0$, depending on the shock speed and local magnetic field \citep{ds96}. 
The enhancement of line emission from shocks and possibly a precursor
region, over and above the emission from gas photoionized by the AGN, 
could then account for the appearance of the bright emission-line knot in Region A associated 
with the radio hotspot.  A moderate contribution to the line emission from
shocks is also consistent with the HST spectroscopic study of \citet{ferruit99}. From the
relative strengths of the [SII] $\lambda\lambda 4069,4077$ and 
$\lambda\lambda 6717,6731$ lines, which are produced profusely in 
the post-shock cooling zone, they estimate high temperatures 
(greater than $20,000-50,000$ K) and somewhat low electron densities (around
$300$ cm$^{-3}$) for the [SII]-emitting gas. These values are quite reasonable
for normal diffuse ISM that has been compressed by a shock front. 
Based on a simple double gaussian decomposition of the H$\beta$ 
line integrated across the knot, we estimate an approximate linewidth 
of $420$ km s$^{-1}$ for the broad component, which includes the high velocity 
wings. This nicely matches the predicted shock velocities from \citet{ferruit99}
which were based on comparisons to shock ionization models of \citet{ds96}.
While it is difficult to directly associate shock speeds with integrated line kinematics,
this broad consistency adds credence to a shock-driven origin for the emission of the
knot in Region A.  


\section{The Nature of the Jet Interaction}

Our combination of high spatial resolution optical spectroscopy and imaging provide the best
existing view of the compact jet-ISM interaction in NGC 5929.  High velocity gas is 
only seen at the location of the south-western radio hotspot, while in other
parts of the NLR, linewidths are narrow and a velocity gradient expected from virial motion
is seen, consistent with ground-based slit and IFU spectroscopy \citep{keel85, stoklasova09}.

We adopt a simple model for the jet interaction, following that of \citet{whittle86}.
The radio jet plows through the ISM as it propagates outward, driving strong shocks
at its head (the radio hotspot). As this shocked gas cools and gets denser, 
it becomes the [\ion{S}{2}]-emitting gas described by \citet{ferruit99} and creates the 
broad H$\beta$ emission. \citet{allen08} model the properties of post-shock gas
as a function of shock velocity $V_{sh}$, pre-shock density and magnetic field strength.
Taking a magnetized shock as fiducial, the [\ion{S}{2}]-derived densities and 
$V_{sh} \sim 400$ \kms (from the linewidth of the broad H$\beta$ component)
imply a pre-shock density of $n \sim 30$ cm$^{-3}$ and a shock cooling length 
around tens of parsecs. The pre-shock density estimate has a large uncertainty
and could be as low as 1 cm$^{-3}$ if the shocks are weakly magentized, 
which would also decrease the cooling length significantly. However, since
Seyfert jets are generally associated with mG level magnetic field, we adopt our estimates for
magnetic shocks as most likely. The cooling length we derive is resolved in our HST images
and may explain the small offset of $0\farcs15$ between the location of the radio hotspot and the 
peak of the line emission in Region A. This interpretation is very tentative, since
patchy dust obscuration is the region may also cause such an offset. 

\citet{su96} estimate flux densities, radio spectral indices and source sizes 
for the various components of the radio source, from which we calculate radio 
source equipartition pressures, using the relations of \citet{miley80} and 
assuming that the radio emitting plasma has a filling factor of unity 
and an ion fraction $a=10$, and that the radio spectrum extends from 
0.01 GHz to 100 GHz with a constant spectral index of -0.82. The equipartition synchrotron 
pressure of the radio hotspot is $10^{-7}$ dyne cm$^{-2}$,
which may be compared to the ram pressure of the shock front, 
$m_{p}\:n\:V_{sh}^2 \sim 8\times10^{-8}$ dyne cm$^{-2}$. The two pressures are 
comparable, consistent with the view that the shocked gas surrounds 
and confines the radio hotspot. It is worthwhile to note that such pressures are 
considerably higher than those typically found in Seyfert radio sources, which may indicate that
this is a relatively young jet ejection event. Eventually, interactions with the
surrounding gas will confine the jet flow into a static lobe, like those found in late-stage
interactions \citep{capetti96, whittle04}.

\section{Discussion: Shock Signatures in Seyferts} \label{sec4}

Why do we see such obvious signatures of shocks in this object and not in others? 
This may be due to the weakness of the AGN: NGC 5929 has an absorption-corrected 
hard X-ray luminosity of $1.8\times 10^{41}$ erg s$^{-1}$ -- low compared to average 
local Seyferts \citep{cardamone07}. Or perhaps the nuclear environment of the 
galaxy is dense and gas-rich due to its ongoing merger. This 
will lower the ionization parameter of nuclear radiation
and produce a more compact emission line region. In both scenarios, the influence of the 
nucleus at larger radii will be relatively unimportant, making shock 
ionized line emission visible against the general background of centrally 
ionized gas. If this is indeed the case, it implies that radiative shock
ionization from nuclear outflows is widespread in Seyfert NLRs, but is usually
secondary to nuclear photoionization processes and only visible
in low-luminosity Seyferts or those with dense nuclear environments.
A rich nuclear environment raises another possibility that the jet is impacting
a dense molecular cloud, enhancing the shock luminosity. This may explain
why no clear shock signatures are seen around the NE radio hotspot, though
the obscuration of the main dust lane prevents a direct view of this
region.

The luminosity of the broad H$\beta$ component is $8\times 10^{38}$ erg s$^{-1}$.
Following \citet{ds96}, the H$\beta$ flux scales with the total radiative flux from the
shock, with a weak dependence on $V_{sh}$, giving shock
luminosities of $5 \times 10^{41}$ erg s$^{-1}$ -- comparable or slightly less than 
the total luminous output of the AGN (taking a X-ray bolometric correction of $\sim 10$).
Using standard relationships from \citet{osterbrock89}, the H$\beta$ luminosity
can be used to derive the mass of ionized gas in the broad component: $3 \times 10^{4}$ M$_{\odot}$.
If this mass of gas was accelerated to $V_{sh}$, it would have a total kinetic energy
of  $5 \times 10^{52}$ ergs. Taking the approximate acceleration timescale to be
the crossing time of the region of size 0\farcs3 (48 pc) at $V_{sh}$ (around $10^{5}$ yr),
a lower limit on the `kinetic luminosity' of the jet is estimated to be $1.5 \times 10^{40}$ erg s$^{-1}$.
This is few to several percent of the total AGN energy output. 
Given that jet outflows are a relatively common feature of Seyfert activity
and couple strongly to the ISM through shocks, jet driven feedback can 
effectively carry as much as energy as the AGN photon luminosity to kpc scales, and 
transfer a tenth or less of this energy in the form of kinetic energy to the NLR. This
can have important consequences for the suppression of bulge star-formation 
and the energy budget and dynamics of circum-nuclear gas.  

\acknowledgments
We thank the referee for their helpful review. DR acknowledges the support of
NSF grants AST-0507483 and AST-0808133. Based on observations made 
with the NASA/ESA Hubble Space Telescope. MERLIN is a National Facility 
operated by the University of Manchester at Jodrell Bank on behalf of STFC.

\end{document}